\documentclass[journal]{vgtc}                

\ifpdf
  \pdfoutput=1\relax                   
  \usepackage{graphicx}                
  \DeclareGraphicsExtensions{.pdf,.png,.jpg,.jpeg} 
\else
  \usepackage{graphicx}                
  \DeclareGraphicsExtensions{.eps}     
\fi%

\graphicspath{{paper/figures/}{figures/}{pictures/}{images/}{./}} 

\usepackage{microtype}                 
\PassOptionsToPackage{warn}{textcomp}  
\usepackage{textcomp}                  
\usepackage{mathptmx}                  
\usepackage{times}                     
\usepackage{cite}                      
\usepackage{tabu}                      
\usepackage{booktabs}                  

\usepackage{xcolor}
\usepackage{xspace}
\usepackage{textcomp}
\usepackage{enumitem}
\usepackage{array}
\usepackage{scalerel}
\usepackage{pbox}

\definecolor{revised}{HTML}{fdc086}

\definecolor{myRed}{HTML}{E75B57}
\definecolor{myBlue}{HTML}{2E65FF}

\definecolor{timeColor}{HTML}{2E65FF}
\definecolor{errorColor}{HTML}{009b76}
\definecolor{evidenceColor}{HTML}{E6550D}

\definecolor{myGreen}{HTML}{009b76}
\newcommand{\task}[1]{{\fontfamily{lmss}\selectfont \small{\textbf{\textcolor{myGreen} {#1}}}}}

\definecolor{myGrey}{HTML}{4d4d4d}

\newcommand{\compoundTask}[1]{{\fontfamily{lmss}\selectfont\small{\textbf{\textcolor{black}{#1}}}}}

\newcommand{\elementaryTask}[1]{{\fontfamily{lmss}\selectfont\small{\textbf{\textcolor{black}{#1}}}}}

\newcommand{\fabio}[1]{{\color{red} Fabio: [{#1}]}}

\newcommand{\highlight}[1]{{#1}}

\definecolor{sanddune}{rgb}{0.59, 0.44, 0.09}
\definecolor{seagreen}{rgb}{0.18, 0.55, 0.34}

\setlist{nosep, noitemsep, leftmargin=*}

\newcommand{\visLabel}[1]{{\fontfamily{cmtt}\selectfont {#1}}}

\newcommand{\myparagraph}[1]{\noindent\textbf{#1.}\xspace}
\newcommand{\hide}[1]{}

\onlineid{1239}

\vgtccategory{Research}
\vgtcpapertype{Evaluation}

\title{A Comparison of Spatiotemporal Visualizations\\ for 3D Urban Analytics}

\author{Roberta Mota, Nivan Ferreira, Julio Daniel Silva, Marius Horga,\\Marcos Lage, Luis Ceferino, Usman Alim, Ehud Sharlin, Fabio Miranda}

\authorfooter{
\item
 Roberta Mota, Julio Daniel Silva, Usman Alim, and Ehud Sharlin are with the University of Calgary, Canada. E-mails: \{roberta.cabralmota, jd.silva, ualim, ehud\} @ucalgary.ca.
\item
 Nivan Ferreira is with the Universidade Federal de Pernambuco, Brazil. E-mail: nivan@cin.ufpe.br.
\item
 Marius Horga and Fabio Miranda are with the University of Illinois at Chicago, United States. E-mails: \{marius, fabiom\} @uic.edu.
 \item
  Marcos Lage is with the Universidade Federal Fluminense, Brazil. E-mail: mlage@ic.uff.br.
  \item
  Luis Ceferino is with the New York University, United States. E-mail: ceferino@nyu.edu.
}

\shortauthortitle{Biv \MakeLowercase{\textit{et al.}}: Global Illumination for Fun and Profit}

\abstract{Recent technological innovations have led to an increase in the availability of 3D urban data, such as shadow, noise, solar potential, and earthquake simulations. 
These spatiotemporal datasets create opportunities for new visualizations to engage experts from different domains to study the dynamic behavior of urban spaces in this under explored dimension. 
However, designing 3D spatiotemporal urban visualizations is challenging, as it requires visual strategies to support analysis of time-varying data referent to the city geometry.
Although different visual strategies have been used in 3D urban visual analytics, the question of how effective these visual designs are at supporting spatiotemporal analysis on building surfaces remains open. 
To investigate this, in this paper we first contribute a series of analytical tasks elicited after interviews with practitioners from three urban domains.
We also contribute a quantitative user study comparing the effectiveness of four representative visual designs used to visualize 3D spatiotemporal urban data: spatial juxtaposition, temporal juxtaposition, linked view, and embedded view.
Participants performed a series of tasks that required them to identify extreme values on building surfaces over time. Tasks varied in granularity for both space and time dimensions. 
Our results demonstrate that participants were more accurate using plot-based visualizations (linked view, embedded view) but faster using color-coded visualizations (spatial juxtaposition, temporal juxtaposition).
Our results also show that, with increasing task complexity, plot-based visualizations perform better in preserving efficiency (time, accuracy) compared to color-coded visualizations.
Based on our findings, we present a set of takeaways with design recommendations for 3D spatiotemporal urban visualizations for researchers and practitioners.
\highlight{Lastly, we report on a series of interviews with four  practitioners, and their feedback and suggestions for further work on the visualizations to support 3D spatiotemporal urban data analysis.}
} 


\CCScatlist{ 
 \CCScat{K.6.1}{Management of Computing and Information Systems}%
{Project and People Management}{Life Cycle};
 \CCScat{K.7.m}{The Computing Profession}{Miscellaneous}{Ethics}
}

\teaser{
  \centering
  \includegraphics[width=\linewidth]{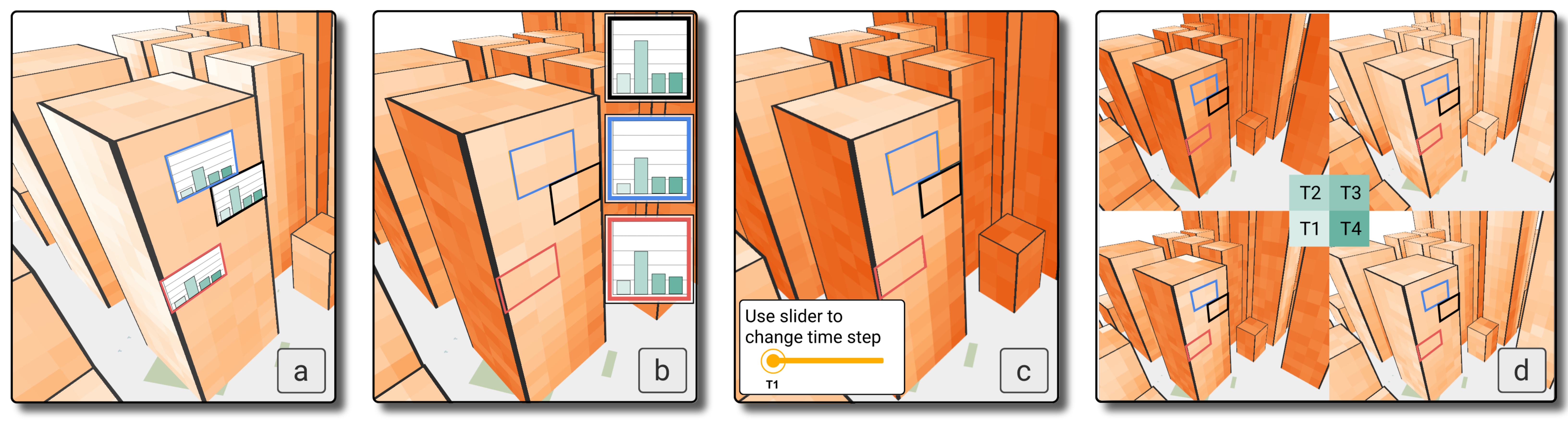}
  \vspace{-0.7cm}
  \caption{The four \highlight{spatiotemporal} visualizations compared in our study: (a) \visLabel{Embedded View}, (b) \visLabel{Linked View}, (c) \visLabel{Temporal Juxtaposition}, and (d) \visLabel{Spatial Juxtaposition}.
   In this example, the visualizations encode the values of an artificially-created temporal attribute \highlight{of the building facade with four time steps.}
  In each visualization, three \highlight{rectangular} regions \highlight{with colored outlines} are placed along the building surface. 
  \highlight{When comparing their attribute values}, the \textcolor{myBlue}{blue} region has the minimum average attribute value between $[t_{1}, t_{2}]$ time steps, whereas the \textcolor{myRed}{red} and black regions have the maximum average values in $t_{3}$ and $t_{4}$, respectively.
  }
	\label{fig:teaser}
}

\vgtcinsertpkg

\begin{document}



\maketitle

\section{Introduction}


Cities are the loci of economic activity and attract people looking for the myriad of services only offered by urban centers.
With the growth of urbanization, the process and problems that shape and form cities become far more intertwined---transportation, housing, street layouts, and land use are all becoming more complex and interconnected by the day.
In their attempt to meet the demands of an increasing number of urban dwellers, cities resort to planning initiatives that lead to the verticalization of their landscape. 
Far from being just another element of cities, their verticality is perhaps the \emph{key} feature that defines cities. 
%
%
While serving as a potential solution to many of the problems faced by cities (\textit{e.g.}, housing~\cite{barr2015skyscraper, barr2020skyscrapers}) verticalization can create new problems or exacerbate existing ones, \textit{e.g.}, their impact on overall temperature~\cite{ipcc6threport}.
%
Cognizant of this phenomenon and the fact that cities are characterized by their verticality, a growing number of domains \highlight{(\textit{e.g.}, civil engineering, urban planning, architecture)} have gone beyond the usual flatland that defines a spatial region, and incorporated the third dimension (and its properties) in many of their analysis tasks in order to study the dynamic behavior of urban spaces. These tasks often rely on data that is intrinsic to the surface of buildings. Some examples include sound propagation simulation for studying noise in the urban environment~\cite{stoter2008}, solar energy potential on facades for the installation of renewable energy equipment~\cite{KAYNAK2018134}, outdoor heat exposure~\cite{Xiaojiang2022}, sunlight access for the design of living spaces~\cite{MORELLO200926, 8283638, Hraska2019}, and cellphone signal propagation through urban canyons for the deployment of cellular towers~\cite{Ahmadien2020}.
%
%
%
%
%
%
%
The transitions to more sustainable environments, energy sources, housing, and technologies have underscored the importance of leveraging the geometry of cities in its entirety. As important in understanding cities as it is, 3D urban analytics has remained fairly unexplored.
Unlike visual analysis on 2D maps, which has been thoroughly explored in the past, visualizations and visual analytics tools designed with 3D capabilities (and more importantly, that enable 3D-oriented tasks) are few and far between.

Designing visualizations \highlight{for domain experts} with 3D urban scenarios in mind is challenging, as it requires visual strategies to support analysis of the data referent to the city geometry.
Tackling this is fundamental to uncovering features that are valuable for decision-making and problem solving in the aforementioned domains.
Our goal in this work is, therefore, twofold. Recognizing the fact that 3D urban analytics is an unexplored yet growing area, our first goal is to have a deeper understanding of the requirements and analysis workflow usually employed by domain experts and practitioners in 3D urban analytics. To achieve this goal, we conducted a series of semi-structured interviews with experts from different domains \highlight{in order to elicit a list of requirements that should be satisfied by visual analytics in the urban environment.}

\highlight{Our second goal is to better understand the effectiveness of different design strategies in the visualization of 3D spatiotemporal urban data}: spatial juxtaposition, temporal juxtaposition, linked view and embedded view. To achieve this goal, based on the previously elicited requirements, we extended these designs to incorporate the geometries of cities and allow the visualization of data on building surfaces. We then performed a quantitative user study where participants performed a series of tasks that required them to identify extreme values on building surfaces over time. Our results point to the most \highlight{performant} designs to solve common 3D-oriented analytical tasks in the urban environment.
Our contributions can be summarized as follows:

\begin{itemize}
\item A \textbf{task characterization} to inform visualization researchers and practitioners new to the 3D urban domain. We report on a series of semi-structured interviews with domain experts that routinely perform analytical tasks taking into account the 3D urban environment, and summarize visualization tasks encountered in this domain.
    
\item A controlled \textbf{quantitative study}  comparing \highlight{user performance} of four representative spatiotemporal visualizations, in which 32 participants were asked to perform tasks that required the identification of \highlight{regions with} extreme \highlight{attribute} values on building surfaces over time. \highlight{We report and interpret study results, leading to a set of design recommendations on visualization choices for performant spatiotemporal analysis in 3D urban data.}

\item \highlight{A series of \textbf{expert interviews} with four practitioners from the urban domain to get their perceptions and ideas for further work on the visualizations to support analysis of 3D time-varying urban data.}
\end{itemize}

\newcolumntype{L}[1]{>{\raggedright\let\newline\\\arraybackslash\hspace{0pt}}m{#1}}
\newcolumntype{C}[1]{>{\centering\let\newline\\\arraybackslash\hspace{0pt}}m{#1}}
\newcolumntype{R}[1]{>{\raggedleft\let\newline\\\arraybackslash\hspace{0pt}}m{#1}}

\section{Related Work}
\label{sec:litReview}

In this section, we review previous visualization designs that combine space, time, and thematic data dimensions.
%
The review is not restricted to urban contributions and includes general visualizations, to allow us to establish a set of common designs. We also survey and classify 3D urban visualizations according to the previously identified designs.

\subsection{Visualization of 2D/3D spatiotemporal data}
\label{sec:litReviewDesigns}
A survey conducted by Andrienko \textit{et al}. on spatiotemporal visualizations for 2D geographic data mentions that existing visual strategies can be grouped into two categories \cite{Andrienko2010}. \textit{Linked views} segregate and display spatial and temporal dimensions in multiple coordinated views. \textit{Embedded views} merge spatial and temporal aspects into a single view.

\noindent\textbf{Linked views} is a standard approach to display time-dependent geospatial data. It benefits from presenting data separately, which may prevent the occlusion of useful map information due to interference between the spatial and temporal data. However, a significant limitation of this approach is the additional screen space required to show side-by-side views. Additionally, the spatial decoupling between the 2D and 3D views may lead to mental burden related to the continuous context switching while relating the spatial and temporal data~\cite{ferreira2013, yang2016, deng2019}.
Turkay \textit{et al}. proposed a visualization system to study how attributes vary over geographical spaces \cite{turkay2014}, where users create a linear selection sequence across a 2D map using a semi-automated brushing. After the brush sequence is computed, multiple linked sparklines are arranged in a 2D table layout, with sparkline depicting single attribute changes.


\noindent\textbf{Embedded views} typically use the map to convey the spatial referencing and superimpose time series plots as glyph overlays to present the changes in the data over time. Liu \textit{et al}. embedded a circular time axis enclosing road trajectories on a map, and time-dependent attributes such as speed and duration of those trajectories are displayed as color-coded circular bar charts \cite{liu2011}. Andrienko \textit{et al}. directly embedded ThemeRiver plots onto geographical maps, but this method led to severe clutter and occlusion of other map information \cite{andrienko2004}. To mitigate occlusion issues, Sun \textit{et al}. applied a non-linear zooming algorithm to broaden map roads and overlaid temporal displays on the roads \cite{sun2016}. Abstraction and aggregation can also serve the purpose of de-cluttering and integrating space and time. Andrienko \textit{et al}. discussed aggregation methods used for movement data \cite{andrienko2008}.
%
%
Kim \textit{et al.} group spatial 4D scientific visualizations into four categories: \emph{juxtaposition}, \emph{superimposition}, \emph{interchangeable}, and \emph{explicit encoding}~\cite{Kim2017}. Gleicher \textit{et al.} follow a similar categorization in their taxonomy for infovis techniques~\cite{gleicher2011}. These typologies are relevant as they discriminate visual strategies based on how data attributes are presented to the user, and as a result \textit{which visual and motor perceptual mechanisms are used while comparing multiple data instances}.
 
\noindent\textbf{Juxtaposition} displays multiple data instances at the same time, but in different coordinate spaces, \textit{e.g.}, side-by-side views ~\cite{coffey2013, lu2008, keefe2009}. The benefit is that all instances are simultaneously visible for comparison so that the user does not need to rely on their memory of past visualizations. However, juxtaposition leads to spatial decoupling, which hinders comparative judgments of slight variations between data instances. \highlight{Additionally, scalability of juxtaposition poses two main challenges: available screen space and computational resources. When the number of data instances is high, the data encoding displayed in multiple views may become too small and hard to read. To deal with limited screen space, strategies include filtering to hide data instances that are of no interest to the user \cite{konev2014watchers}, clustering \cite{lu2008}, or abstracting the visual representation of the data to take less screen area \cite{keefe2009}. In order to maintain interactive frame rates as several model instances are being concurrently rendered---progressive computation and sampling can be used to alleviate computational costs \cite{johnson2019, choo2013}.}



Johnson \textit{et al}. propose multiple volumetric views juxtaposed in a grid arrangement for exploratory analysis of moderate-sized 4D cardiac data ensembles on the order of 10 data instances~\cite{johnson2019}. Each grid column refers to a single data instance, \textit{i.e.}, one simulation run from the ensemble at a given time step. An entire grid row of volumetric views is created whenever users specify a box-shaped selection of a sub-volume of interest on a given view (grid cell). In doing so, the technique follows a ``global-to-local'' exploratory approach: users progressively select sub-volumes of interest to interactively build up a comparative spatially-juxtaposed view layout; users can make visual comparisons of key features of interest across data instances and/or time steps.



\noindent\textbf{Interchangeable} displays only one data instance at a time and gradually transitions over time through either an automated or user-controlled animation; the data instances are co-registered in the same coordinate space. This has the advantage that each data instance is displayed at a full scale, without occlusion from other instances. However, since all data instances to compare are not simultaneously visible on the screen, users must rely on remembering past instances. This cognitive load increases with an increasing number of instances  and may hinder comparison, especially for more detailed analyses~\cite{woodring2006, ribivcic2012, waser2014}. Akiba~\textit{et~al.} proposed an animation tool for volume visualization where users create a sequence by combining various individual effects that are set using transfer functions~\cite{akiba2009}.



\noindent\textbf{Superimposition} displays two or more data instances at the same time and coordinate space. This is the only approach that displays multiple data instances in a way that is spatially co-registered, simultaneously visible, and preserves the original data. However, clutter and occlusion are major challenges that superimposed visual designs must overcome to allow the comparison between multiple data instances. 
Glyphs are commonly used for overlaying various data instances in a 3D space~\cite{sanyal2010,rocha2016,zhang2015}. Van Pelt \textit{et al}. proposed a technique that combines established scientific visualizations with novel glyphs whose designs are based on concepts from information visualization \cite{van2014}. The glyph-based visualizations support interactive filtering and details-on-demand zooming to set different zoom levels and provide progressively richer information within the glyph.
%

\noindent\textbf{Explicit encoding} does not display multiple data instances, but rather derives a composite data from all the original instances and only displays the composite result (\textit{e.g.}, difference). The advantage of this approach is that the derived composite is displayed in full scale. In addition, for well-defined tasks, it is possible to set tailored encodings and visualize the composite result directly. On the other hand, explicit encoding does not preserve the original data; Therefore it is challenging to make comparisons between individual instances, even when the encoding is known~\cite{frey2012, alabi2012, hao2015}. 
Melanie \textit{et al.} proposed a glyph-based visualization to study the progression of multiple sclerosis lesions from MRI scans over multiple time steps~\cite{tory2001}. 



\begin{figure}[t!]
    \centering
    \includegraphics[width=0.7\linewidth]{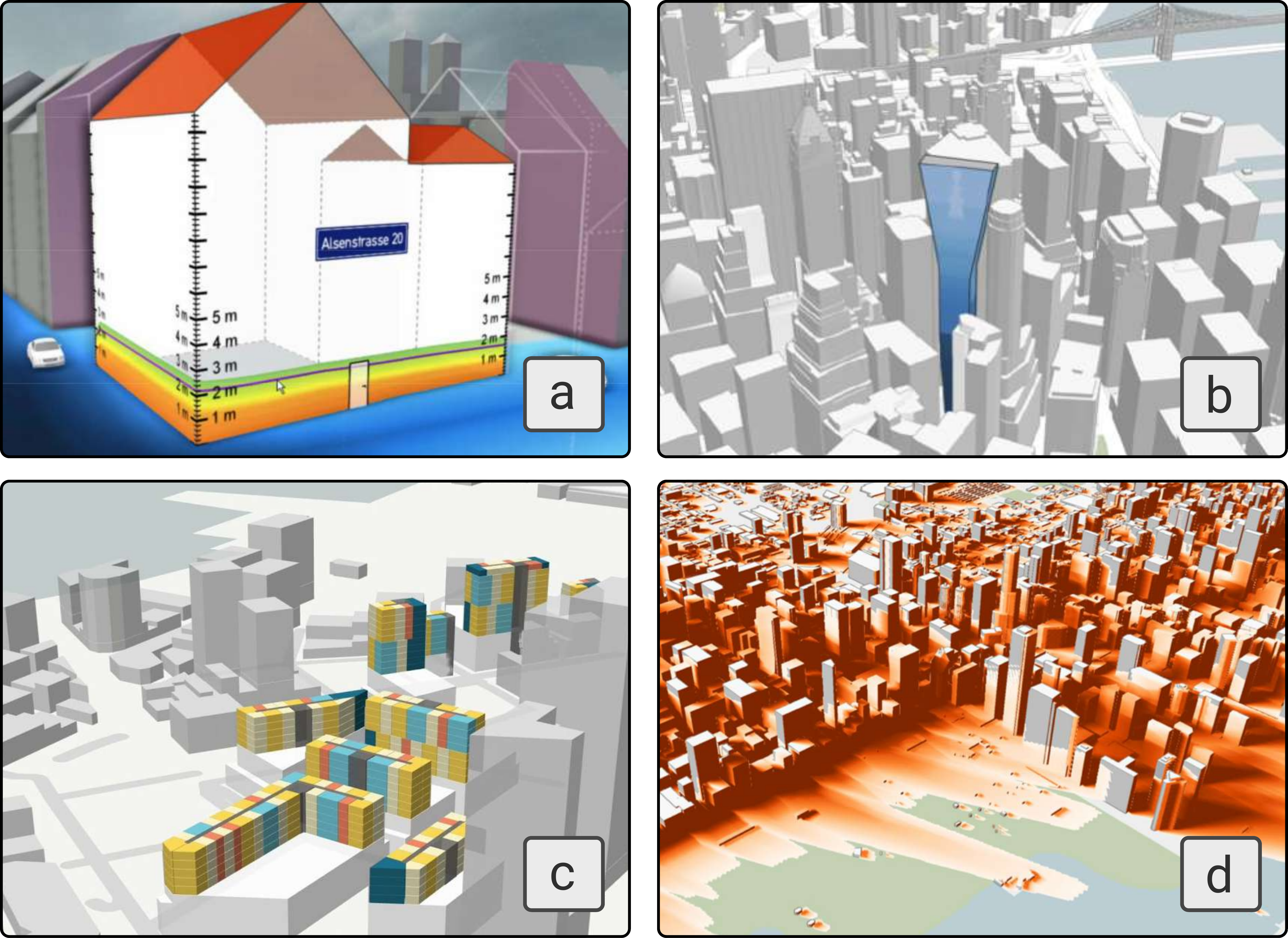}
    \vspace{-0.3cm}
    \caption{Examples of data on urban surfaces from different domains: (a) flooding simulation analysis in the environmental domain \cite{vuckovic2021}, (b) view impact analysis in urban planning  \cite{doraiswamy2015}, (c) building design performance analysis in urban design \cite{Delve}, and (d) solar potential and sunlight access analysis in urban planning and architecture \cite{8283638}.}
    \label{fig:urbanSystems}
    \vspace{-0.5cm}
\end{figure}

\subsection{Visualization of 3D spatiotemporal urban data}
%
Visualization systems are important tools for the analysis of urban data~\cite{zheng2016visual,8474495}.
The majority of these applications use a flat city metaphor---\textit{i.e.}, a 2D map, to represent the city environment\cite{ferreira2013,zheng2016visual,Doraiswamy:2018:IVE:3183713.3193559,karduni2017urban,10.1145/3313831.3376399}.
However, cities have inherent geometric features, leading to an increase in the number of 3D urban visualization systems in recent years \highlight{(see Figure \ref{fig:urbanSystems})}. 
%
These systems justify the need for 3D since specialists evaluate and communicate to stakeholders using 3D urban models (\textit{e.g.}, discussing building designs with varied impacts on shadows, landmark visibility, and sky exposure~\cite{8283638, Urbane, doraiswamy2015, Vis-A-Ware}). 


After learning about the various visual designs for spatiotemporal data, we wanted to obtain an overview of the designs commonly used for 3D urban data.
For this purpose, we reviewed recent 3D urban visualization systems that deal with multi-dimensional 3D urban data and classified them according to the visual designs discussed in the prior section---see classifications in Table \ref{tab:tableClassificationVis}.
It is worth noting that, at a high level, the elicited spatiotemporal visual designs are essentially comparative designs that can be used for any multi-dimensional data type if reinterpreted slightly for the situation where multiple data dimensions are compared. 
Hence, our review includes all multi-dimensional data types such as temporal, multivariate, and uncertain.

From the literature, we searched papers on publication channels that publish quality work on visual analytics. These venues were either from the visualization or urban domains---\textit{e.g.}, IEEE Transactions on Visualization and Computer Graphics; Computers \& Graphics; Computer Graphics Forum; and Symposium on Simulation for Architecture and Urban Design. For these venues, although not exhaustively, we surveyed the archive published over the last ten years and selected \highlight{34} papers presenting 3D urban visualization systems.
Table \ref{tab:tableClassificationVis} displays a classification of recent 3D urban visualization systems. We observe that, concerning visual designs coming from 3D sci-vis, 3D urban vis systems heavily rely on \textit{interchangeable}~(\highlight{27}), followed by \textit{juxtaposition}~(7), and very few systems seem to use \textit{explicit encoding} (2) and \textit{superimposition} (1). Regarding designs drawn from 2D info-vis, we observe that most 3D urban vis systems do include 2D abstract visualizations (\highlight{19}). Among these, \textit{linked views} is considerably more commonly used than \textit{embedded view} (12 and \highlight{7}, respectively).


%

%



%

\begin{table}[t!]
  \begin{center}
  \caption{Classification of 3D urban visualization systems. The rows group systems according to Kim \textit{et al.}'s taxonomy for 3D spatiotemporal scientific visualizations~\cite{Kim2017}. The first two columns group according to Andrienko \textit{et al.}'s topology for 2D spatiotemporal information visualizations~\cite{Andrienko2010}. The third column refers to 3D urban vis systems that do not include 2D visualizations. 
  }
  \vspace{-0.2cm}
  \centering
    \begin{tabular}{ m{0.32\linewidth} || m{0.12\linewidth} | C{0.17\linewidth} | C{0.18\linewidth} }
       & \visLabel{LinkedV} & \visLabel{EmbeddedV} & \visLabel{No 2D}\\
       \hline\hline
      \visLabel{Juxtaposition} & \small \cite{wilson2019, konev2014watchers, Vis-A-Ware, waser2011, goldstein2020, krietemeyer2019} &  &  \small \cite{noyman2020}    \\
      \hline
      \visLabel{Interchangeable} & \small \cite{zeng2018vitalvizor, Urbane, gautier2020, waser2014, waser2011, rockcastle2018} & \small \cite{Zhang2021urbanvr, vuckovic2021, cornel2015, sicat2018, ribivcic2012, cornel2016, trapp2019}  &  \small \cite{xu2021, kilsedar2020, 8283638, cornel2019, konev201825, ramyar2019, ribicic2012, malm2019, schaumann2019, christodoulou2018, dogan2018, konig2021, semmo2015, pasewaldt2013}  \\
      \hline
      \visLabel{Superimposition} & \small \centering \cite{Vis-A-Ware} &  &    \\
      \hline
      \visLabel{Explicit encoding} & \small \centering \cite{ribivcic2012}  &  &  \small \cite{8283638, wolosiuk2020, konig2021}\\
    \end{tabular}
    \vspace{0.25cm}
    \vspace{-1cm}
    \label{tab:tableClassificationVis}
  \end{center}
\end{table}

\section{Study Rationale}
\label{sec:Analyses}

%
%
Unlike well-established sub-areas of visualization, such as scientific visualization, 3D urban visualization relies on ad-hoc solutions, primarily driven by industry standards or one-off collaborations between visualization researchers and domain experts.
Our goal in this work is then to offer a more comprehensive understanding and characterization of \highlight{visualization} tasks and visualization \highlight{designs} that support exploration of 3D time-varying urban data.
In order to achieve this, we structure our efforts in two different phases. First, we review the literature and perform a series of interviews with domain experts to elicit representative tasks and visualizations used in the urban domain.
Second, guided by our previous findings, we select a subset of the elicited tasks and design a quantitative user study to measure the effectiveness of four techniques for the visualization of 3D spatiotemporal urban data.

This section reports on the first phase of our work.
In Section~\ref{sec:task}, we detail interviews with experts and outline common tasks they perform for time-varying urban data analysis in 3D city models.
%
In Section~\ref{sec:vis}, we discuss representative approaches to visualizing 3D time-dependent urban data, with the second phase of our work detailed in Section~\ref{sec:study}.
\subsection{Task characterization}
\label{sec:task}

We conducted a series of one-hour semi-structured interviews with domain experts: a civil engineer interested in urban resilience, disaster management, and earthquake simulations; an urban planner actively involved in the analysis of the solar potential of buildings; and an architect interested in the design of public and living spaces with more equitable access to sunlight and views to landmarks.
Our first aim during these interviews was to better understand the types of temporal analyses they commonly perform with the purpose of deriving a \emph{task characterization} that covers a range of representative tasks that 3D spatiotemporal urban visualizations must be able to support. 
%
%
Since these visualizations combine spatial, temporal, and thematic data dimensions, we frame our characterization after these data facets as four guiding \textbf{questions} asked during interviews (Q1-4), followed by a \textbf{discussion} of common use cases in their domains (D1-4), and condensed \textbf{responses}~(R1-4).




\noindent \textbf{Q1: {What actions are relevant for temporal data analysis in 3D city models?}}\\
D1: In 3D city models, the visual analyses of spatiotemporal attributes are driven mainly by the spatial rather than the temporal dimension. Actions include \emph{spatial search} (\emph{e.g.}, search for buildings with increasing solar incidence throughout the day), \emph{spatial comparison} (\emph{e.g.}, compare shadow behaviors between different facades; which one is most shadowed during the winter?), and \emph{temporal comparison} (\emph{e.g.}, compare shadow behavior between summer and winter months on a given street; in which season does the shadow distribution have a sharper increase? In which month does it receive less shadow?).\\
R1: \task{Spatial search}, \task{Spatial comparison}, \task{Temporal comparison}

\noindent \textbf{Q2: {What spatial granularities are relevant for temporal data analysis in 3D city models?}}\\
D2: Domain specialists may be interested in different granularity levels. Architects run performance-driven analysis of candidate building designs (\emph{e.g.}, daylight performance due to its influence on aesthetics and perception of space, energy conservation, and occupants' health and satisfaction). The evaluation of candidate designs typically operates on the \emph{facade} level, and this includes both the \emph{entire} and custom \emph{region} of the facade. Whereas urban planners commonly perform multi-scale spatial analysis: from the \emph{street} to \emph{facade} to \emph{building} to the \emph{neighborhood} level~(\emph{e.g.}, how does the shadow behavior on a given \emph{street}/\emph{facade}/\emph{building} vary due to a new building development?).\\
R2: \task{Street}, \task{Facade region}, \task{Facade}, \task{Building}, \task{Neighborhood}

\noindent \textbf{Q3: {What temporal granularities are relevant for temporal data analysis in 3D city models?}} \\
D3: Typical temporal resolutions include individual \emph{time steps} and custom \emph{time intervals}, and this may vary on a case-by-case basis according to the analysts' needs. For instance, when examining wind circulation behavior on a given neighborhood, they can focus their analyses on peak hours, or \emph{monthly} and  \emph{seasonal} (\emph{e.g.}, summer, winter) trends.\\
R3: \task{Time step}, \task{Time interval}, \task{Monthly}, \task{Seasonal}

\noindent \textbf{Q4: {What attribute behaviors are relevant for temporal data analysis in 3D city models?}}\\
D4: Domain experts commonly examine \emph{extreme} attribute values (\emph{e.g.}, which street receives most shadow during summer?), \emph{average} values (\emph{e.g.}, on a given facade region, what is the average solar incidence during winter months?), and \emph{trends} (\emph{e.g.}, on a given facade, is there  small variation in wind circulation throughout the day?).\\
R4: \task{Extremes}, \task{Averages}, \task{Trends}

In our task characterization, the combination of the previous dimensions allow us to derive visualization tasks. For example, a common visualization task reported in urban planning involves the evaluation of sunlight access in the buildings of a given neighborhood, which can be specified as a \task{spatial search} for \task{facades} with \task{extreme} attribute values at certain \task{months} of the year. We build on this characterization to elaborate representative tasks for our user study (Section~\ref{sec:study}).


\subsection{Visualization analysis}
\label{sec:vis}




During our interviews, we learned that our domain collaborators worked with time-varying scalar functions defined on grids on top of the building surfaces.
This data type therefore is set as the focus of study in this paper.
We also learned that the experts' approach to visually examining time-dependent 3D urban data consists of color-coding attributes on the building surfaces, and using manual/automatic animations to display the data over time. 
They mentioned that due to the 3D nature and complexity of urban surfaces, relevant information may be missed or occluded and therefore requires users to restart the animation several times to inspect patterns in the visualization.
%

%
Following the interviews, we reviewed the literature in order to find alternative visual designs used to integrate temporal information into their \highlight{3D} spatial references. 
As described in Section~\ref{sec:litReview}, our literature review identified a number of possible designs from which we selected four for comparison in our user study.

\subsubsection{Selection of visualizations}

To make our study tractable, our criteria for selecting representative visualization designs was twofold. 
First, we chose designs based on familiar 2D information visualizations in order to assess the efficiency of two ways of combining the physical data (building geometry) and thematic (temporal) data: \textit{linked} and \textit{embedded views}.
Second, concerning designs based on 3D scientific visualizations, we chose only those derived from \textit{juxtaposition} and \textit{interchangeable} approaches. 
Not only are these familiar to the domain specialists, they are also the most popular designs used for 3D spatiotemporal data visualizations~\cite{Kim2017}.

%
%


Our decision to discard explicit encoding designs is due to the fact that they do not present the entire original data values but rather only composite values derived from the original instances. 
Visualizing a meaningful derived relation between two or more data instances may provide a more focused encoding than would be possible by visualizing all instances separately. 
However, we are interested in designs that allow users to visualize individual data instances and to judge variations between them to maintain accordance with experts' analysis workflow.

Moreover, we also discard techniques based on superimposition since they often suffer from overplotting and strongly rely on carefully-tailored encodings~\cite{Kim2017}. 
The superimposed visual encodings need to not only communicate data semantics, but also be perceptually separable from one another so that users can perceive each encoding individually (\emph{e.g.}, understand and discern encodings referent to a specific time interval).
Therefore, we discard superimposition as we feel they would become a design challenge that could possibly bias study results.



Based on our choices, we identify four representative visual designs used to combine thematic, spatial, and temporal data into one visual representation. The visualizations can be classified into two major groups according to the graphical encoding of the spatiotemporal thematic attribute: \emph{plot-based} and \emph{color-coded} designs.
Plot-based designs draw from 2D information visualizations and visually encode the attribute as the \emph{position on a common scale} of an abstract time series plot.
Color-coded designs build on 3D scientific visualizations and map the attribute to the \textit{color} visual channel of the building geometry itself. 
The four visual designs are rationalized below, taking into account the fact that the data type of our time-varying thematic attribute is quantitative in nature. We refer the reader to the supplementary video for dynamic frames of each visualization.

\begin{figure}[t!]
    \centering
    \includegraphics[width=0.9\linewidth]{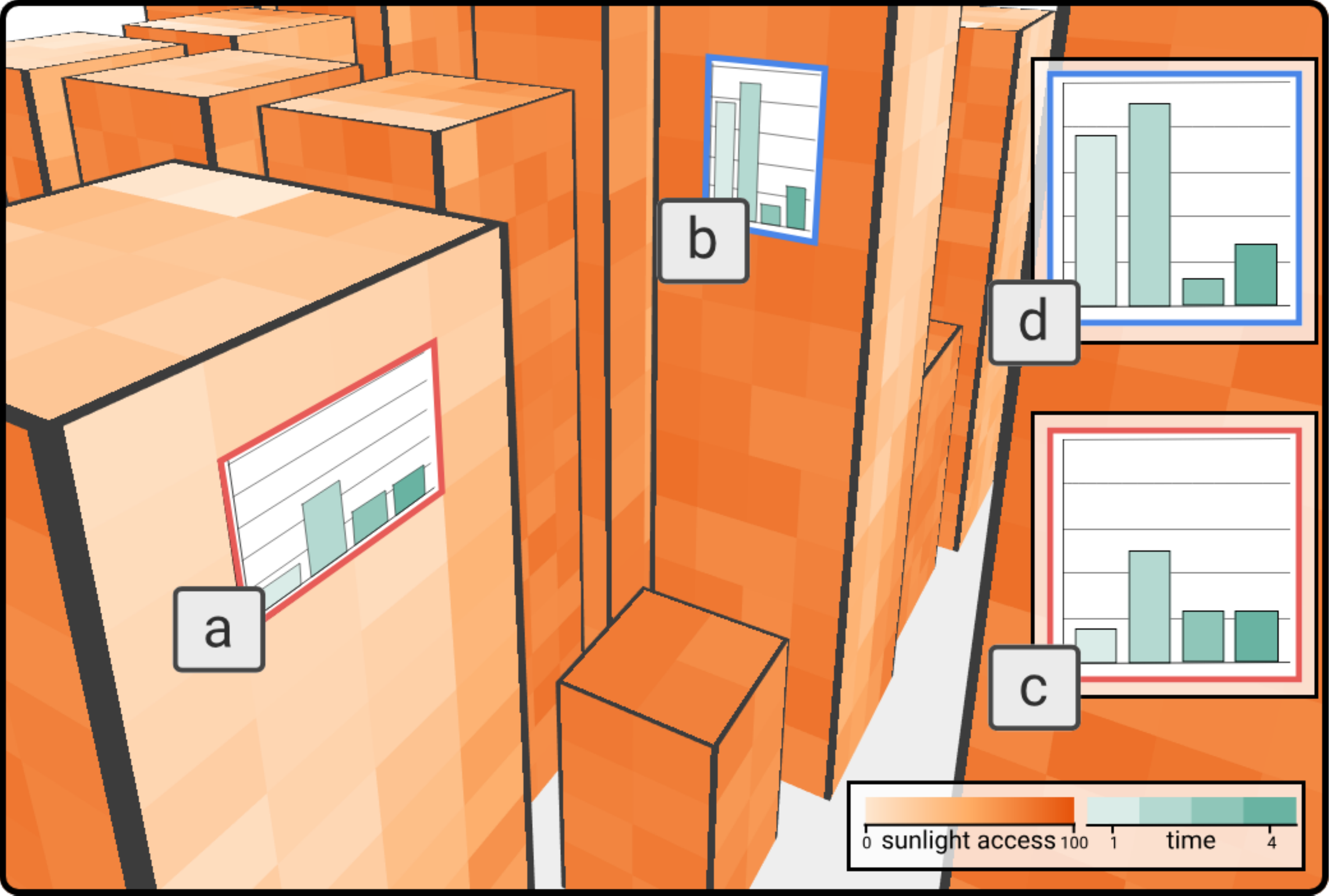}
    \vspace{-0.3cm}
    \caption{Reference lines may assist comparative data analysis in at least two ways \highlight{for \visLabel{EmbeddedV}}. (a, b) First, when plots are embedded in building geometries with discrepant resolutions. Second, as perspective may distort visual assets. A line, as well as the white area formed between the top of a bar and that line, provides common baseline scales for data comparison--\textit{e.g.}, when comparing values in time $t4$. (c, d) \highlight{Direct view of the same plots displayed in (a), (b) as a comparative baseline of perceptual artifacts:} the same plots without perspective distortion and with equal resolution.}
    \vspace{-0.2cm}
    \label{fig:refLines}
\end{figure}

\noindent \textbf{Embedded view} \highlight{(\visLabel{EmbeddedV})} displays all data instances on a bar chart embedded in the building surface geometry---Figure~\ref{fig:teaser}(a).
The quantitative attribute values are encoded as the bar charts' ordinates, while time stamps are encoded as abscissas. The choice for encoding an attribute value as the vertical position on a common scale (bar chart) is because this visual channel is known to be one of the most effective for accurately decoding information~\cite{munzner2014}. 
%
Furthermore, adding a coordinate system together with reference lines can facilitate the judgment and comparison of different bars. This may be useful to mitigate perceptual issues due to perspective distortion and discrepant aspect ratios of different buildings---Figure~\ref{fig:refLines}(a,b).

To embed the bar chart into the building surface, we use a decal-based approach similar to Rocha \textit{et al.}~\cite{rocha2016}. A decal map is built from the intersection of a sphere with a building surface; that is, the part of the surface that lies inside the sphere. This leads to an area that follows the surface geometry, like a decal. Afterwards, a local parametrization is computed to map the bar chart texture to the building's geometry.

\noindent \textbf{Linked view} \highlight{(\visLabel{LinkedV})} displays all data instances on a bar chart similar to embedded view; however, the chart is arranged on the right side of the screen, as depicted in Figure \ref{fig:teaser}(b).

\noindent \textbf{Temporal juxtaposition} \highlight{(\visLabel{TemporalJX})} displays a single data instance at a time in full scale, and a user-controlled slider widget allows switching to other instances---see Figure~\ref{fig:teaser}(c). A sequential colormap encodes the temporal attribute, and this same coloring is used throughout all time steps.

\noindent \textbf{Spatial juxtaposition} \highlight{(\visLabel{SpatialJX})} arranges all data instances in an evenly-spaced $2 \times 2$ grid layout---see Figure \ref{fig:teaser}(d). Since data instances are not registered in the same coordinate space,
we extend the views with linked cameras and linked region highlights in order to provide a common context between the multiple views. A sequential colormap encodes the time-dependent quantitative attribute (\emph{e.g.}, shadow).

\highlight{These visualization designs, while satisfying our previously mentioned criteria, also posses certain drawbacks: \visLabel{EmbeddedV} can distort visual information~\cite{ware2019}; \visLabel{LinkedV} requires additional screen space for side-by-side views and the decoupling between 2D and 3D views may lead to extraneous cognitive load related to the continuous context switching~\cite{ferreira2013, yang2016, deng2019}; \visLabel{TemporalJX} does not show all data instances simultaneously, users must rely on memory of past instances which also increases cognitive load~\cite{woodring2006, ribivcic2012, waser2014}; and \visLabel{SpatialJX} requires available screen space (see Section \ref{sec:litReviewDesigns} for more details).}



\subsubsection{Interactions}
The selected visualization designs support interactivity. The user can control the camera with the mouse---rotation via the right mouse button, zoom using the mouse wheel, and panning by pressing and holding the mouse wheel while moving the mouse.
Furthermore, in each visualization, users can also interact with regions. We define a \emph{region} as a compound area composed of $2 \times 2$ grid cells along the building surface. The \emph{region's attribute value} refers to the average values of all cells composing that region.  \highlight{In this sense, the color-coded visualizations (\visLabel{SpatialJX}, \visLabel{TemporalJX}) require users to mentally compute the mean value within a region based on its cells' colors. In the plot-based visualizations (\visLabel{LinkedV}, \visLabel{EmbeddedV}), users judge a region's mean value based on the vertical position along the bar chart.}

Participants create a region by clicking the left mouse button over the building surface; drag a region by pressing and holding the middle mouse button over it while moving the mouse; and delete a region by clicking on it with the right mouse button. Users can define up to four regions simultaneously to compare \highlight{attribute values} on different parts of the building surface; regions are discriminated by color---\highlight{Figure~\ref{fig:teaser} depicts three regions in each visualization design}.

\section{Study Design}
\label{sec:study}
We now present our user study, whose goal was to compare the effectiveness of the four visualization techniques discussed in Section~\ref{sec:vis}.
The study followed a between-subjects design---participants were only exposed to one of the visualization techniques---and simulated an analysis scenario in which users were asked to perform some of the visualization tasks described in Section~\ref{sec:task}.
Participants performed a total of 12 tasks. The task presentation order was counterbalanced using Latin squares, and the visualization assignment for the participants was randomized.  In total, the study consisted of 32 participants $\times$ 12 tasks = 384 trials.

\myparagraph{Datasets} We synthetically generated two 3D city models---one model for training and another for the main trials. The models incorporated common urban patterns (\textit{e.g.}, medium-density downtown), and had, on average, 100 buildings with \highlight{regular shapes and varying sizes}. For each model, the time-varying attribute defined on the facades was created by discretizing each facade into $5\textsf{m}\times5\textsf{m}$ cells. For each cell, we computed the accumulated sunlight access throughout 8 hours of a day for each one of the four yearly seasons, each season referring to one timestamp. Lastly, we added noise to increase data variability.



\myparagraph{Apparatus} The visualizations were implemented using Unity. The experiments were conducted remotely on an Intel Core i3 and GeForce GTX Titan and used Zoom and Parsec for remote access.


\begin{figure}[t!]
    \centering
    \includegraphics[width=0.9\linewidth]{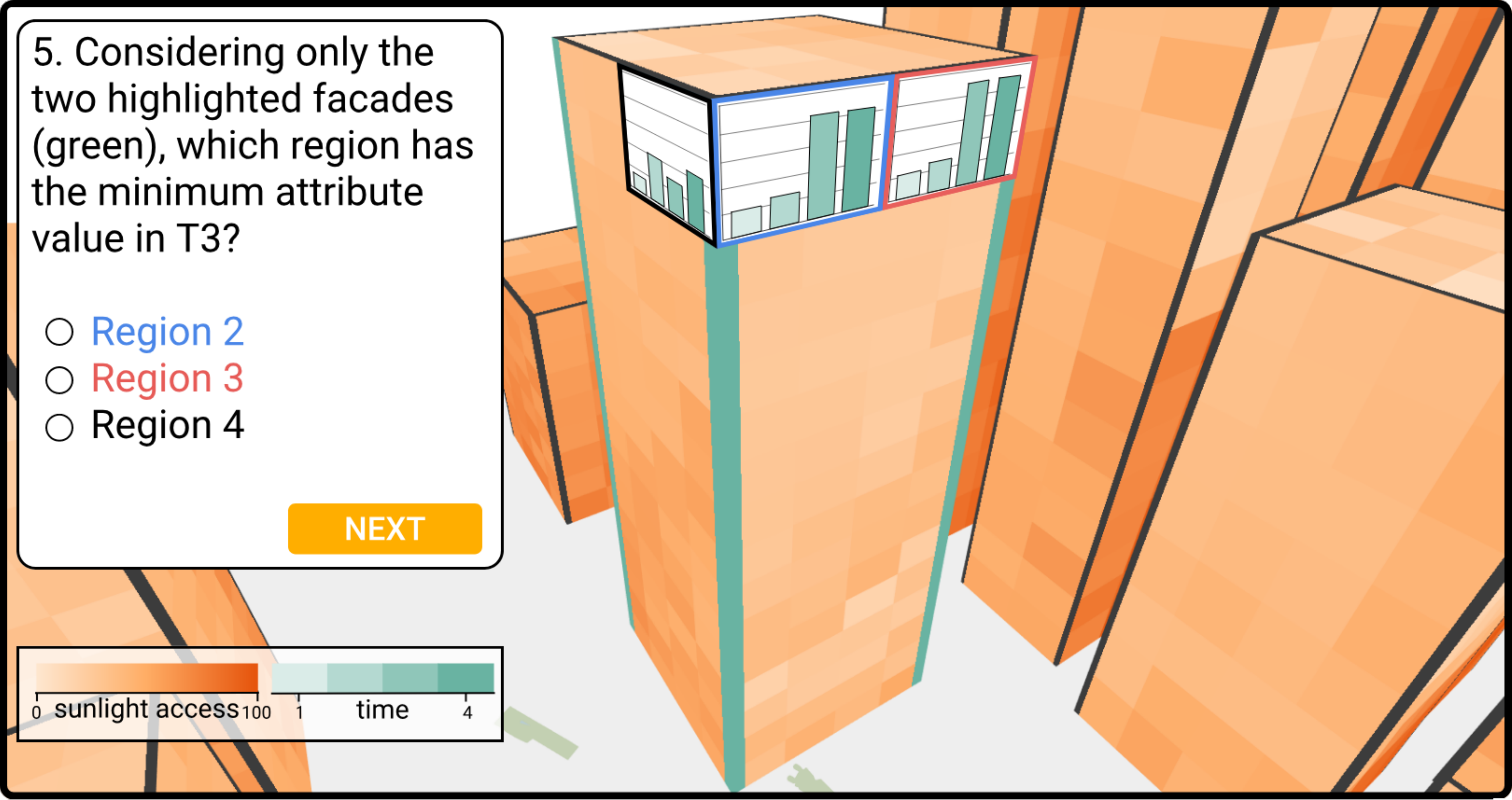}
    \vspace{-0.3cm}
    \caption{ \highlight{\visLabel{EmbeddedV} trial (Task \#5, \elementaryTask{multi-orientation 1 location} in \elementaryTask{1 timestep}),} with three user-created regions: \textcolor{myBlue}{blue} (region 2), \textcolor{myRed}{red} (3), and black (4).}
    \vspace{-0.5cm}
    \label{fig:interface}
\end{figure}


\begin{figure*}[t!]
    \centering
    \includegraphics[width=1.0\linewidth]{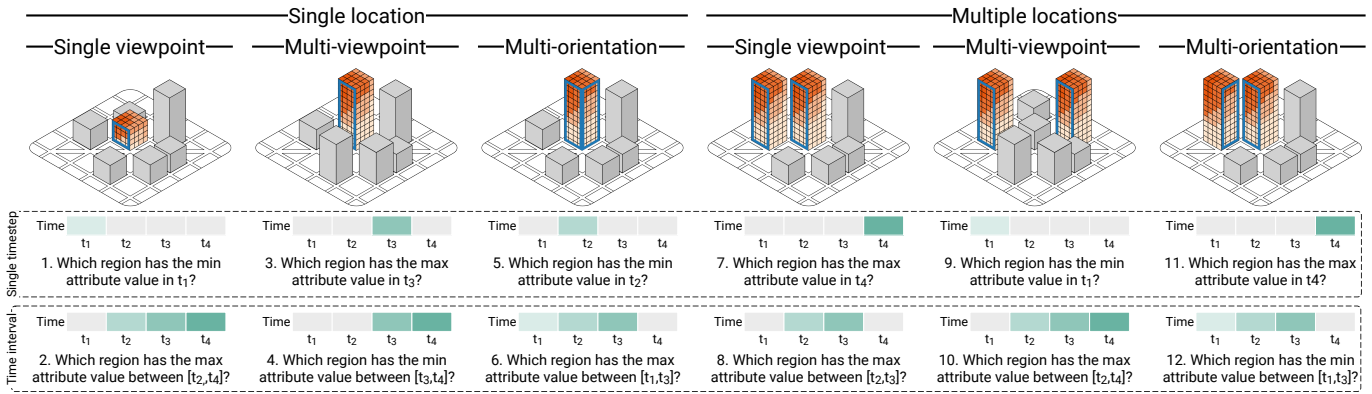}
    \vspace{-0.5cm}
    \caption{Illustrative scheme of the twelve tasks of our study, with varying levels of spatial (\textit{where}) and temporal (\textit{when}) granularities. In each cell, the image illustrates the task scenario with highlighted facade(s), together with a sample question (\textit{what}).}
    \vspace{-0.5cm}
    \label{fig:study_conditions}
\end{figure*}

\myparagraph{Participants} We recruited a total of 35 participants but had to exclude one participant who declared during the session that s/he had misunderstood how to perform the tasks. We also excluded participants who had internet connection issues during the remote session.
From the final 32 participants (9 female and 23 male), none reported any color deficiency, and they had normal or corrected-to-normal vision. Age ranged from 21 to 47 ($\textrm{avg.}= 29.18$, $\textrm{std. dev.} = 6.38$), and most of them were students (26 out of 32) from Computer Science graduate programs. Their backgrounds were mainly in visualization and HCI. They were all volunteers, and did not receive any monetary compensation.
%

\subsection{Procedure}

Each session consisted of three parts: introduction and tutorial, main trials, and post-study questionnaire (60--90 minutes total).
In the introduction and tutorial, we explained the purpose of the experiment and the visualization to be used in the session. Then, the participants answered five training trials. After each trial, we would indicate if the answer was correct or not. They were encouraged to ask questions throughout the tutorial. If they made no errors and declared that they had no further questions, they would start the main trials.
In the main trials, participants answered 12 questions and received no feedback regarding the correctness of their answers. They were asked to answer as quickly as possible without compromising accuracy. 

\highlight{Figure 4 shows a trial screenshot. For each trial, the building facade(s) pertaining to the trial was highlighted in green. Users could create, delete, or move  regions along these target highlighted facade(s). On the top-left corner, the text box depicts the task number and question for the trial. It also lists all existing regions at any given time. By selecting a region on the list and clicking the `next' button, the user submits that region as the task answer. On the right is the 3D urban visualization; the same visualization was used across the training and main trials---\visLabel{SpatialJX}, \visLabel{TemporalJX}, \visLabel{LinkedV}, or \visLabel{EmbeddedV}.} 

\subsection{Measures}
For each trial, we defined two objective metrics: 1) \textbf{task completion time} is measured from the moment participants saw the trial screen until they submitted an answer; and 2) \textbf{relative error} is computed as the absolute difference between the chosen attribute value and the true value, normalized by the task's maximum possible absolute error.
During each trial, we also logged qualitative observations about participants' behaviors, interaction strategies, and aloud comments. After completing the trial, participants also filled out a post-study questionnaire to share positive and negative opinions, as well as ideas to improve the visualization used in their session.

\subsection{Tasks}
%

Guided by our task characterization, our study aims to compare the four visualizations in regards to a relevant task: \highlight{to spatially search for the facade region with extreme attribute value over a certain time period. 
We had no hypothesis about which action would be more difficult: to characterize a region's attribute value (by averaging its cell values using position or color encoding), or to discriminate a region's attribute variation over time.
Hence, we treated them as an integrated task that requires users to assess both the mean attribute value within a region and its changes over time}, and varied the combinations of these two spatial and temporal dimensions during the construction of our tasks.

Figure~\ref{fig:study_conditions} illustrates our task construction. To define our tasks, we used Peuquet's framework for geo-temporal data that describes the linked triad of \textit{what}, \textit{where}, and \textit{when}~\cite{peuquet1994}. Each task refers to a question in the format: \textit{when} + \textit{where} $\rightarrow$ \textit{what}, where \textit{what} is the participant's characterization of extreme attribute values. We varied the \textit{when} and \textit{where} dimensions in a way similar to other research~\cite{goodwin2015, schiewe2018, pena2019}, using varying spatial and temporal granularity levels. The temporal granularity \textit{when} was divided into \elementaryTask{1 \highlight{timestep}} and \elementaryTask{time interval} (two or three consecutive time steps). The spatial granularity \textit{where} was segmented into \elementaryTask{1 location} (one building), and \elementaryTask{N locations} (two buildings). The spatial granularity was further broken down into:

1. \elementaryTask{Single} and \elementaryTask{multi-viewpoint}: 
\highlight{
\elementaryTask{Single-viewpoint} tasks target a single facade in \elementaryTask{1 location}, and two facades in nearby buildings in \elementaryTask{N locations}. 
\elementaryTask{Multi-viewpoint} tasks target a facade partially occluded by another building in \elementaryTask{1 location}, and two facades in fairly distant buildings in \elementaryTask{N locations}.
For \elementaryTask{single-viewpoint} tasks, users can look at the target highlighted facade(s) from a single point of view. For \elementaryTask{multi-viewpoint} tasks, users cannot capture the entire target facade(s) from a single viewpoint; thus, users must navigate the camera and continuously switch spatial context while searching for a facade region with extreme temporal value.
}

2. \elementaryTask{Single} and \elementaryTask{multi-orientation}:
\highlight{\elementaryTask{Single orientation} tasks target two \elementaryTask{single-viewpoint} facades facing the same direction.
\elementaryTask{Multi-orientation} tasks target two \elementaryTask{single-viewpoint} facades whose surfaces are pairwise orthogonal.}
Note that these conditions occur in both \elementaryTask{1 location} and \elementaryTask{N locations}.

Crossing the spatial and temporal dimensions resulted in a matrix of 12 possible spatiotemporal tasks illustrated in Figure \ref{fig:study_conditions}, along with a \highlight{sample} task \highlight{question} for each one of them.

\begin{table}[b!]
  \begin{center}
  \caption{Description of task conditions (\textit{rows}). Each task condition represents the difference in time and error between the two groups: lower and higher complexity (\textit{columns}).}
  \vspace{-0.5cm}
  \centering
    \begin{tabular}{ m{0.35\linewidth} || m{0.25\linewidth} | m{0.25\linewidth} }
       & \compoundTask{Lower} & \compoundTask{Higher} \\
       \hline\hline
      \compoundTask{Number of time instances} & \elementaryTask{1 timestep} &  \elementaryTask{Time interval} \\
      \hline
      \compoundTask{Number of spatial instances} & \elementaryTask{1 location} &   \elementaryTask{N locations} \\
      \hline
      \compoundTask{Camera navigation} & \elementaryTask{Single viewpoint} &  \elementaryTask{Multi-viewpoint}  \\
      \hline
      \compoundTask{View angle} & \elementaryTask{Single orientation}   &  \elementaryTask{Multi-orientation} \\
    \end{tabular}
    \label{tab:tableTaskConditions}
  \end{center}
\end{table}

\subsection{Task conditions and complexity}
\label{ssec:task_conditions}

%
Although complexity may come in many forms, in accordance with prior research~\cite{gleicher2017}, four conditions that drive the difficulty of our visualization tasks are: \compoundTask{number of time instances}, \compoundTask{number of spatial instances}, \compoundTask{camera navigation}, and \compoundTask{view angle}.
The idea is that an increase in complexity will increase the number of spatial and temporal searches or camera movements that users must perform in order to find the answer.

While being unequivocal that increasing either the \compoundTask{number of time instances} or the \compoundTask{number of spatial instances} leads to an increase in complexity due to the extra searches an expert will be required to perform, \compoundTask{camera navigation} and the selection of a proper \compoundTask{view angle}, in particular, also increase complexity and lead to loss of performance. These depend on implicitly solving the inverse problem of where to direct our view, given a mental image of the rendering we \emph{expect} to see.
This is an inherently ill-posed problem as the expectation inferred from the rendering might not properly correlate with the 3D geometry because of common rendering artifacts (such as distortions caused by the camera projection, occlusion, etc.).
This is typically, and intuitively, solved by trial and error: camera movements are tuned iteratively until adequate choices are found. 
As a concrete example, consider the case where data is presented in multiple locations. Occlusion prevents users from setting a single camera view that would work for all the data, which requires frequent manipulations of the viewpoint each time a different  location is inspected.
%
It is in this way that \compoundTask{camera navigation} and \compoundTask{view angle} conditions impact user performance: due to continuous camera view adjustments and context switching, time and attention are not used towards the user's utmost goal that is to understand the data.

To assess the complexity impact on user performance, we derive four task conditions from the elementary tasks listed in Figure~\ref{fig:study_conditions}. 
For each condition, we select a subset of tasks and divide them into two groups: lower and higher complexity. 
Next, we compute the differences in time and error between the two groups (lower and higher complexity) to obtain the impact in efficiency for that task condition.
Table \ref{tab:tableTaskConditions} illustrates the selection and grouping of elementary tasks for each task condition.

\section{Results}
\label{sec:results}
We now report and interpret findings for task \textbf{completion time} and \textbf{relative error} from a total of 384 trials. We refer the reader to the supplementary material for the data files.
%
Our reporting methodology uses estimation techniques and reports sample means with confidence intervals (CI) rather than $p$-value statistics~\cite{cumming2005, dragicevic2016}, following recommended practices~\cite{cumming2014, american2019}. However, a $p$-value approach can always be obtained from the data by following similar techniques. 
For our inferential analysis, we use pairwise differences between means and their 95\% CIs \footnote{A CI of \textit{differences} that does not cross zero provides evidence of differences. The further away from zero and the smaller the CI, the stronger the evidence.\label{footnote}}, indicating the range of plausible values for the population mean. We use BCa bootstrapping to construct the confidence intervals.






We first provide an overview of the results across the tasks to evaluate the efficiency of all four visualizations (Section~\ref{sec:pairwise}). Afterwards, we present results for varied task conditions to investigate efficiency with increasing task complexity (Section~\ref{sec:dataFactors}).

%


\subsection{Tasks}
\label{sec:pairwise}

\begin{figure}[t!]
    \centering
    \includegraphics[width=0.9\linewidth]{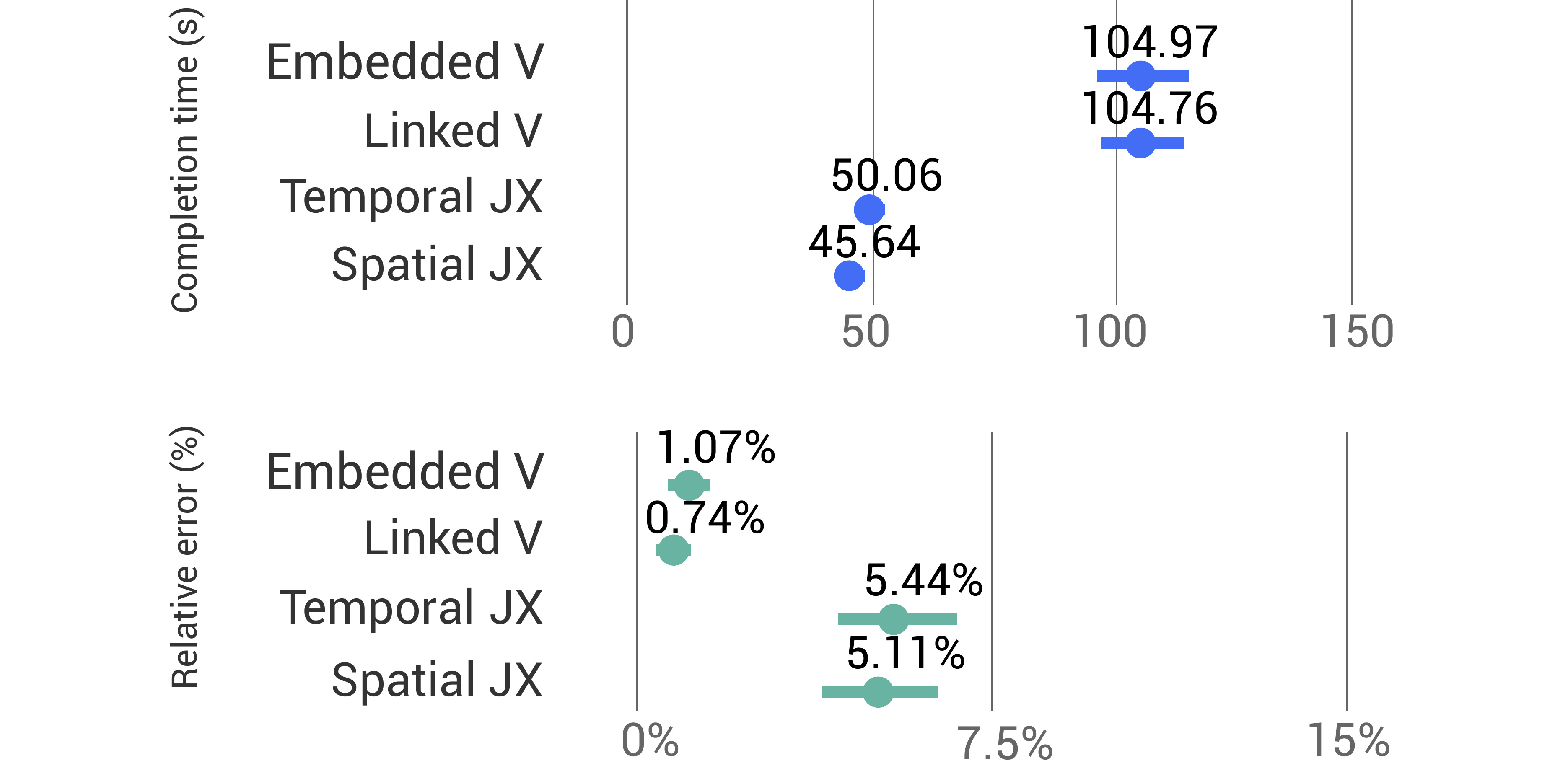}
    \vspace{-0.4cm}
    \caption{ Mean \textcolor{timeColor}{completion time} (\highlight{s}) and \textcolor{errorColor}{relative error} (\%) for each visualization, for all tasks.  Error bars represent 95\% Bootstrap CIs.}
    \vspace{-0.2cm}
    \label{fig:study_TimeErrorAllTasks}
\end{figure}

\begin{figure}[t!]
    \centering
    \includegraphics[width=0.9\linewidth]{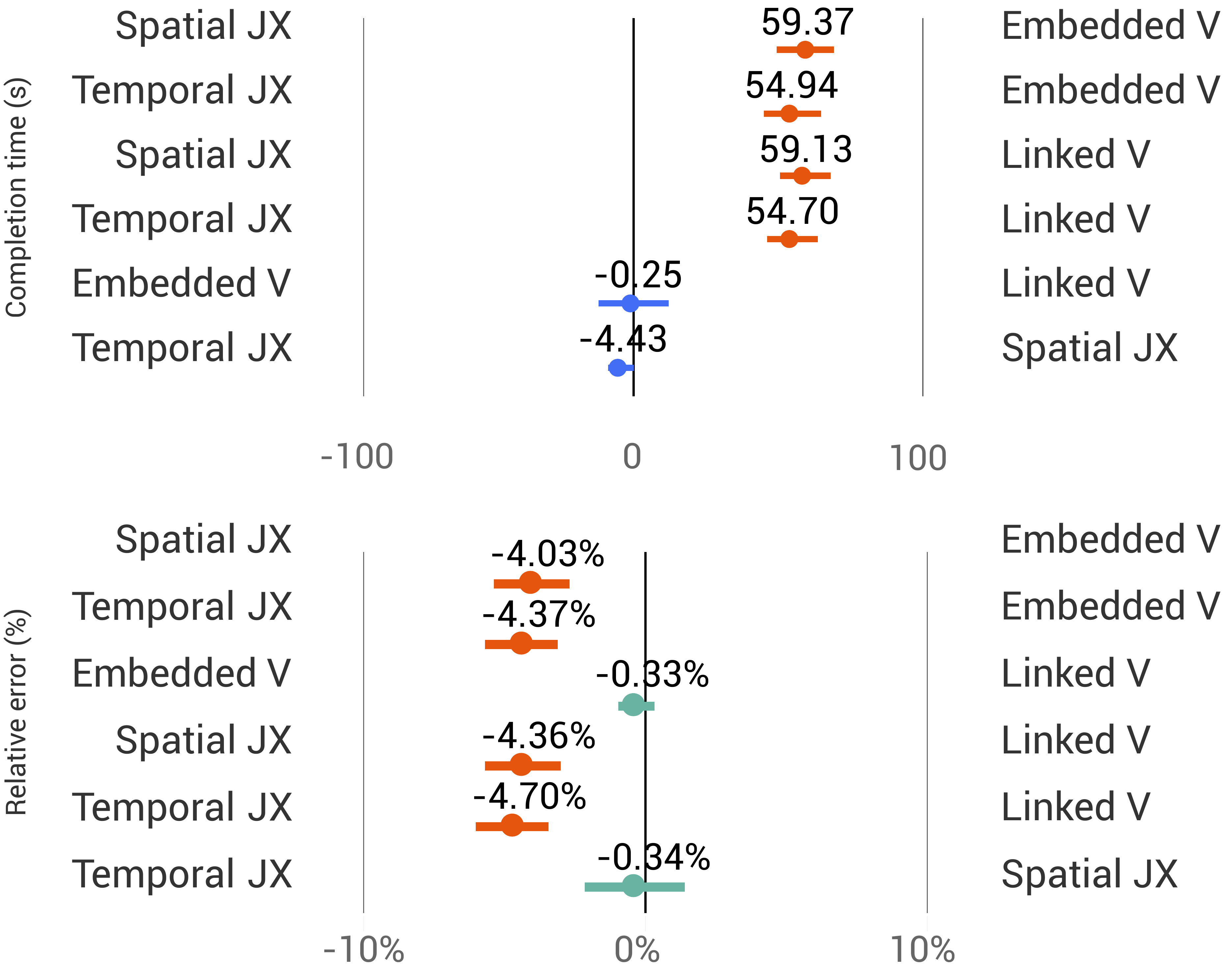}
    \vspace{-0.4cm}
    \caption{Pairwise comparisons between visualizations for \textcolor{timeColor}{completion time} (\highlight{s}) and \textcolor{errorColor}{relative error} (\%), for all tasks.   Error bars represent 95\% Bootstrap CIs, with evidence of differences in \textcolor{evidenceColor}{orange}.\textsuperscript{\ref{footnote}}}
    \label{fig:study_PairwiseTimeErrorAllTasks}
    \vspace{-0.5cm}
\end{figure}

In this section, we investigate the question: ``is there a difference in efficiency (time, error) between the four visual designs?''. 
Figure~\ref{fig:study_TimeErrorAllTasks}\,(top) displays mean times per visualization and Figure~\ref{fig:study_PairwiseTimeErrorAllTasks}\,(top) shows pairwise time comparisons for all tasks collectively. 
%
%
Participants spent 45.64 seconds on average (CI = [42.93, 48.44]) using \visLabel{SpatialJX}, and slightly longer using \visLabel{TemporalJX} (50.06\highlight{s}, [47.33, 52.85]). There is a pronounced increase in task duration in the results of both \visLabel{LinkedV} (104.76\highlight{s}, [96.36, 113.28])  and \visLabel{EmbeddedV} (104.97\highlight{s}, [95.98, 114.05]).
Pairwise comparisons show strong evidence that plot-based visualizations were slower than color-coded visualizations (by 57.03\highlight{s} on average, [47.78, 66.15]). However, there is no clear evidence that completion times were different within each visualization group---\emph{i.e.}, \visLabel{LinkedV} \textit{vs}. \visLabel{EmbeddedV}, and \visLabel{SpatialJX} \textit{vs}. \visLabel{TemporalJX}.

Figure~\ref{fig:study_TimeErrorAllTasks}\,(bottom) displays mean errors and  Figure~\ref{fig:study_PairwiseTimeErrorAllTasks}\,(bottom) shows mean error differences for all tasks collectively.
Participants' average error was 1.07\% ([0.62\%, 1.57\%]) when using \visLabel{EmbeddedV}, and 0.74\% ([0.38\%, 1.15\%]) with \visLabel{LinkedV}. When using color-coded juxtaposed visualizations, the average error increased to 5.11\% ([3.89\%, 6.37\%]) with \visLabel{SpatialJX} and to 5.44\% ([4.21\%, 6.72\%]) with \visLabel{TemporalJX}.
In this case, there are significant differences between plot-based and color-coded visualizations: the former is 4.36\% on average ([3.06\%, 5.69\%]) more accurate than the latter. However, there is no clear difference within each visualization group.

\subsection{Task conditions and complexity}
\label{sec:dataFactors}
In addition to differences in efficiency between the visualizations, we were also interested in investigating if their efficiency is affected with increasing task complexity, \emph{i.e.}, ``for each visual design, is there a difference in efficiency (time, error) with an increasing task complexity?''.  This sections reports results of efficiency impact based on the four task complexity conditions discussed in Section \ref{ssec:task_conditions}---namely, \compoundTask{number of time instances}, \compoundTask{number of spatial instances}, \compoundTask{camera navigation}, and \compoundTask{view angle}.

\begin{figure*}[t!]
    \centering
    \includegraphics[width=0.9\textwidth,keepaspectratio=true]{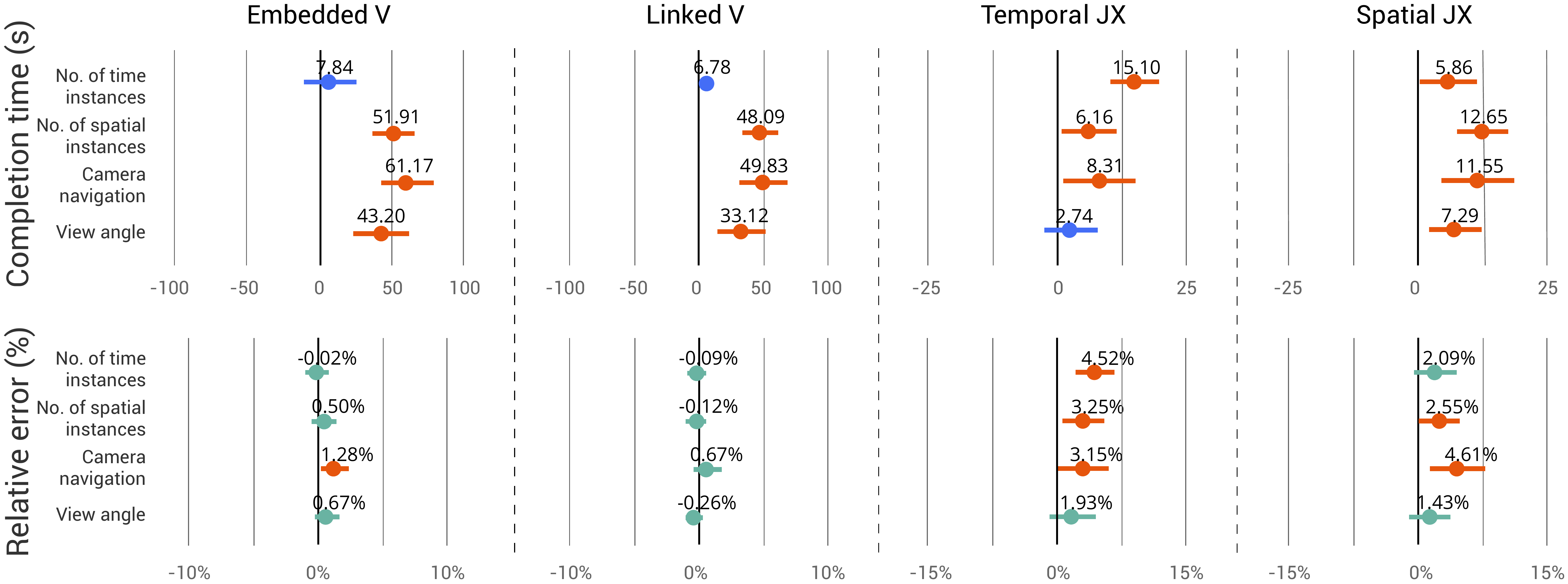}
    \vspace{-0.3cm}
    \caption{Increase in \textcolor{timeColor}{completion time} (\textit{\highlight{s}}) and \textcolor{errorColor}{relative error} (\textit{\%}). Rows: task conditions. Columns: visualization designs. \highlight{Note that plots may present different time/error scales.} Error bars represent 95\% Bootstrap CIs, with evidence of differences in \textcolor{evidenceColor}{orange}.\textsuperscript{\ref{footnote}}}
    \label{fig:study_ResultsPerCondition}
    \vspace{-0.5cm}
\end{figure*}

\subsubsection{Completion time}

Figure~\ref{fig:study_ResultsPerCondition}\,(top) shows results across the task conditions and design. For the \visLabel{TemporalJX} visualization, the \compoundTask{number of time instances} had the most impact on increasing completion time (15\highlight{s}, [10.4, 19.77]), followed by \compoundTask{camera navigation} (6.31\highlight{s}, [2.28, 15.3]) and \compoundTask{number of spatial instances} (6.16\highlight{s}, [0.81 , 11.62]). However, comparison within the \compoundTask{view angle} condition shows no clear increase in completion time.
For \visLabel{SpatialJX}, the \compoundTask{number of spatial instances} and \compoundTask{camera navigation} had the most impact on task completion time: 12.65\highlight{s} ([7.65, 17.64]) and 11.54\highlight{s} ([4.64, 18.76]), respectively; this dropped for \compoundTask{view angle} (7.29\highlight{s}, [2.22, 12.35]) and \compoundTask{number of time instances} (5.86\highlight{s}, [0.47, 11.39]).
For \visLabel{LinkedV}, the \compoundTask{number of spatial instances} and \compoundTask{camera navigation} had a significant impact on increasing task duration: 48.09\highlight{s} ([33.96, 61.59]) and 49.83\highlight{s} ([30.87, 68.35]), respectively. This dropped to 33.12\highlight{s} ([13.85, 51.68]) in the \compoundTask{view angle} condition, but no significant time difference was found with an increasing \compoundTask{number of time instances}.
For \visLabel{EmbeddedV}, similar to \visLabel{LinkedV}, the \compoundTask{number of spatial instances} and \compoundTask{camera navigation} had a strong impact on increasing task duration: 51.9\highlight{s} ([36.88, 66.6]) and 61.16\highlight{s} ([42.87, 78.96]). This dropped to 43.2\highlight{s} ([23.85, 62.06]) in the \compoundTask{view angle} case. There is no evidence that increasing the \compoundTask{number of time instances} led to an increase in completion time.


\subsubsection{Relative error}
Figure~\ref{fig:study_ResultsPerCondition}\,(bottom) shows the results on task error increase across the four task conditions.
For \visLabel{TemporalJX}, \compoundTask{number of time instances} had the most significant impact on task error increase (4.52\%, [2.2\%, 6.82\%]), followed by  \compoundTask{number of spatial instances} (3.25\%, [0.82\%, 5.66\%]) and \compoundTask{camera navigation} (3.15\%, [0.12\%, 6.14\%]).
For \visLabel{SpatialJX}, \compoundTask{camera navigation} had the most significant impact on error (4.61\% increase, [1.44\%, 7.79\%]), and dropped to 2.55\% ([0.06\%, 4.96\%]) with an increasing \compoundTask{number of spatial instances}.
When considering \visLabel{LinkedV}, none of the four task conditions had a significant impact on task error increase; and for \visLabel{EmbeddedV}, \compoundTask{camera navigation} was the only condition that significantly impacted task error (1.28\%, [0.28\%, 2.4\%]).

\section{Discussion} 
\label{sec:discussion}
In addition to examining the quantitative differences in performance between the four visualization designs, we also observed participants' behaviors and strategies when using each visualization. These observations, along with insights from participants' questionnaires, allowed us to more comprehensively characterize how they used each visualization.

\subsection{Tasks}

For the \textit{elementary} tasks, we saw clear differences between plot-based and color-coded visualizations, in terms of both accuracy and task completion time---see Figure \ref{fig:study_PairwiseTimeErrorAllTasks}. 
Participants were considerably more accurate on average when using plot-based visualizations than when using color-coded juxtaposed visualizations. This tendency is consistent with participants' perceived difficulty in conducting the tasks. It was stated often in the post-study questionnaire that  it is hard to perceive minor variability in the data through color encoding. 

\textit{\textbf{Takeaway \#1:}} \textit{Plots produce the most accurate results, and there is no clear winner between linked and embedded visualizations.}

However, the accuracy gain seen in the plot-based designs may have come at the expense of a substantial decrease in overall speed. Participants were considerably slower, possibly due to three factors: 1) plot-based visualizations required users to identify the target attribute values by scanning regions across the entire spatial domain, 2) since plots allowed participants to extract more subtle differences in the data, they likely spent more time on the task verifying their choices, and 3) occlusion---while most highlighted facades on the 3D city model were directly accessible, we observed that partially-occluded areas could make interactions that require direct access difficult (\textit{e.g.}, mouse clicking and dragging regions). As a result, participants in our study often needed to adjust viewing angles (sometimes repeteadly) to reach a certain part of the building, drag the region, and see the visual plot.


\textit{\textbf{Takeaway \#2:}} \textit{Plot-based visualizations require spatial scanning, which makes them slower to identify, verify, and access target data values compared to color-coded visualizations.}

\subsection{Task conditions and complexity}

\subsubsection{Completion time}
For the task conditions, we observed that task completion time increases substantially given an increasing \compoundTask{number of spatial instances}, \compoundTask{camera navigation}, and diverging \compoundTask{view angles} when using plot-based visualizations, possibly because of the three factors previously stated. However, we found no evidence that completion time increases with a higher \compoundTask{number of time instances}. \highlight{This is likely because} plot-based visualizations make it easy to see trends over time: collates all time steps across the individual bar chart; therefore, the cognitive load (time) did not significantly add~up. 
The same was not true for spatially and temporally juxtaposed visualizations, where participants spent longer with an increasing \compoundTask{number of time instances} likely because participants had a fair amount of difficulty to collate information across different spatial and temporal contexts, as we observed. Although spatially-juxtaposed visualization displays all data instances at the same time, collating information across somewhat distant screen areas seems to affect cognitive load (time).

\textit{\textbf{Takeaway \#3:}} \textit{Plot-based visualizations scale better in completion time given an increasing number of time instances as they collate all information needed at a specific location.}


We also observed that the time increase for plot-based visualizations tends to be more than twice the increase for the color-coded ones for task conditions that cover a wider or more convoluted spatial domain---\textit{e.g.}, the \compoundTask{number of spatial instances} condition considers a higher number of building facades, and the \compoundTask{camera navigation} condition deals with partially-occluded facades. 
To illustrate, with an increasing \compoundTask{number of spatial instances}, the time increase for \visLabel{SpatialJX} was 12\highlight{s} against 48\highlight{s} for \visLabel{LinkedV}. This tendency may reflect our prior finding that spatial scanning is a bottleneck for task duration in plot-based visualizations: it requires not only sequential scan but also physical accessibility to partially-occluded surface areas (\textit{\textbf{Takeaway \#2}}).





\begin{figure}[t!]
    \centering
    \includegraphics[width=0.47\textwidth,keepaspectratio=true]{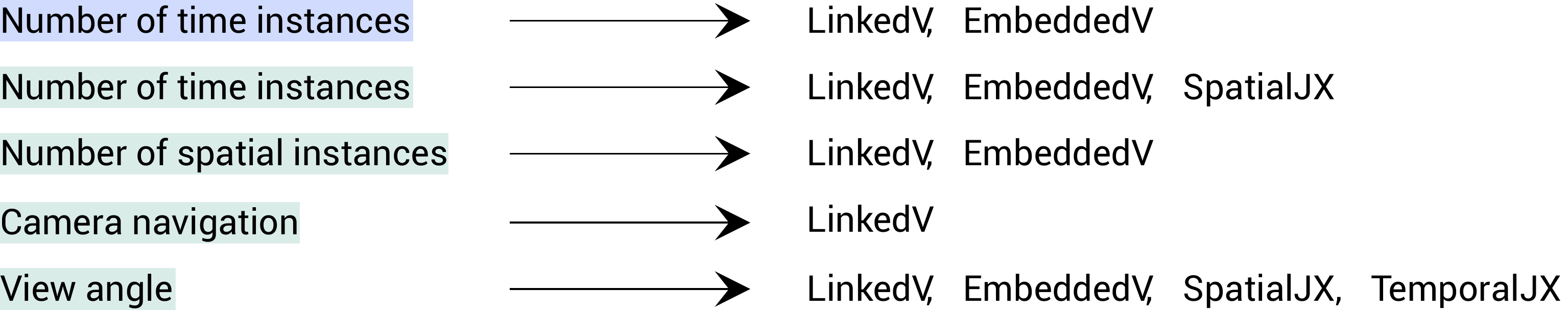}
    \vspace{-0.3cm}
    \caption{The designs that scale best in \textcolor{timeColor}{completion time} and \textcolor{errorColor}{error} with increasing condition complexity (\textit{i.e.}, no evidence of increase in time/error).}
    \vspace{-0.5cm}
    \label{fig:study_Recommendations}
\end{figure}

The results of our study indicate that participants spent substantially longer when using plot-based visualizations; however, these designs scale well when analysis grows difficult with an increasing \compoundTask{number of time instances}, likely because the plots collate all the information needed to identify temporal tendencies. Given these findings, Figure~\ref{fig:study_Recommendations}\,(top) shows recommendations of visual designs that were most effective in task duration with increasing task condition complexity.

\subsubsection{Relative error}

\noindent {\fontfamily{cmtt}\selectfont LinkedV} did not have significant error increase for \textit{any} of the four task conditions. A similar pattern was observed in \visLabel{EmbeddedV} as little evidence of error increase was found only in the \compoundTask{camera navigation} condition---and we believe that the required context switches between the different building facades was the dominant causal factor. Hence, the plot representation seems to sustain accuracy irrespective of the number of locations and timestamps taken into account.

\textit{\textbf{Takeaway \#4:}} {\textit{ Plot-based visualizations scale better in accuracy given an increasing number of spatial instances, temporal instances, or camera navigation. Yet, continuous context switching may be a concern for embedded visualizations.}}

One interesting finding from this study is the inconclusive evidence of task error increase with varying \compoundTask{view angle} for all four visualizations. This is particularly intriguing as \visLabel{EmbeddedV} raises concerns due to distortion of visual information partly caused by perspective. The lack of observed differences may be due to low statistical power; but we believe it is more likely due to the referencing lines added to the plot design. Participants' comments indicate that they often used the lines, as well as the empty space between the top of the bar and the line, to make comparisons within and between individual plots---see Figure \ref{fig:refLines}. Yet, since aliasing artifacts appear on the reference lines with increasing camera distance, we observed that participants had to move close enough to the building model in order to make them legible.


\textit{\textbf{Takeaway \#5:}} \textit{Reference lines can serve as a visual aid in plot-based visualizations to facilitate comparison, at least from a legible distance.}

The results of our study indicate that plot-based visualizations (\visLabel{EmbeddedV} and \visLabel{LinkedV}) can not only enhance the legibility of spatiotemporal data on 3D city models, but also scale well when analysis grows more complex. Based on our observations, Figure~\ref{fig:study_Recommendations}~(bottom) suggests the visual designs that were most effective in task error with increasing task condition complexity.
%
%
Lastly, our design recommendations depicted in Figure~\ref{fig:study_Recommendations} demonstrate that, considering increases in complexity in task conditions, plot-based visualizations more often preserve efficiency (time, accuracy) compared to color-coded ones.




\section{Practitioners' Perspective}
\label{sec:qualitative}

\highlight{In addition to examining quantitative differences in performance between the visual designs, we conducted interviews with four practitioners from the urban domain; two were the same who advised us on task requirements for spatiotemporal analysis in 3D urban visualizations.

\myparagraph{Procedure} Each session was divided into three parts. First, we started with a description of the motivation and goal of our study. Second, we performed four tutorials; these were the same tutorials previously used in our study with the purpose of introducing the analysis tasks and each of the four visualizations.
Throughout the tutorials, participants were encouraged to pose questions and asked to think aloud to express reasoning.
The researcher guided the trials, and adopted an inquiry-based learning approach throughout the tutorial by posing questions to the participants---\textit{e.g.}, ``\textit{can you tell me a location to place a region?}'' and ``\textit{can you indicate which region has the highest average attribute value in $t_{1}$ and $t_{2}$?}'', and relying on them to assess the underlying task analysis and recommend courses of action to complete each trial.
The third part of the session consisted of a semi-structured interview to collect participants’ impressions as well as suggestions for further work on the visual designs.
Sessions lasted 45 to 60 minutes and were video recorded. Comments were transcribed, and an open coding approach was applied. Quotes have been lightly edited to remove extraneous or repeated words, without changing the semantics of the sentence.

\myparagraph{Results} Overall, all participants agreed on the usefulness of the visualizations to support analysis of 3D time-varying urban data.
Although color encoding was acknowledged to be most familiar, three of the four participants reported a preference for the plot representation. They highlighted the fact that plots can inform various features of the temporal function (\textit{e.g}., local/global extrema, increasing/decreasing intervals) that would otherwise be difficult to extract from colors (P1, P2). They also argued that they were more confident when comparing variations between values using plots than they were with colors (P1, P2, P4). One participant reported a preference for a hybrid visualization as ``\textit{colors could be used for rapid screening, and plots could be used to get and compare regions based on more precise data}'' (P3). 
Some participants remarked that embedded views minimize the seam between the plots and their spatial references in a way that reduces mental workload as ``\textit{you no longer need to associate spatial regions with plots (...) you’re (spatially) manipulating the plots themselves}'' (P4). Others remarked that linked views allow plots to be always ``accessible'' (P2) and ``adjacent'' (P1) to one another, even when comparing distant regions.

We also asked participants to provide suggestions for further work to build on the current designs.
Participants suggested investigating plot designs. An example are designs to aid comparison between regions: ``stacked'' (P1) or ``multiseries'' (P3) bar charts, or even charts that encode operations between data points from two or more regions---\textit{e.g.}, ``\textit{the difference in the solar irradiance from candidate areas to install solar panels}'' (P3). The comparative designs could ``\textit{be particularly useful for comparing data from dispersed regions using \textit{e.g.}, embedded views}'' (P1).
Another example is multivariate designs to aid the co-analysis of damages incurred to both outer and inner layers of the building---``\textit{it could be very interesting to see, \textit{e.g.,} in a story, not only the damage (\textit{e.g.}, stress, displacement, fatalities, cost of repair) of the facade but also the damage profile of internal areas (\textit{e.g.}, partition walls) (...) maybe by projecting it to the facade itself}'' (P1).

Participants also suggested considering plot interactions. An example is filtering as ``\textit{time responses of structural damages tend to vary significantly along the building facade due to vertical earthquake motions (...) apply filters on the plots to specify separate time periods makes sense for assessing damage severity \textit{e.g.,} at a story by story level}'' (P3). Other examples include hovering, reordering, and even orienting plots since ``\textit{for vertical earthquake motions, comparing regions’ responses at different story levels (..) `rotating' the bar charts and (reference) lines would likely make it easier comparing them}'' (P3).
Lastly, participants suggested other region shapes to cover areas considered during damage analysis: story, multi-story, and corners (P1, P2, P3).
}
\section{Conclusion, Limitations, and Future Work}
\label{sec:conclusion}

In this paper, we first presented a domain task characterization as a series of abstract tasks that benefit from 3D spatiotemporal urban visualization by conducting interviews with experts from urban planning, architecture, and civil engineering; we hope that this serves as a resource for the visualization community to better understand the growing area of 3D urban analytics.
%
%
%
%
%
Afterwards, we selected a subset of these tasks and presented a quantitative study comparing \highlight{user performance} of four representative visual designs used for the analysis of time-varying 3D urban data. \highlight{Our findings indicate that participants were more accurate using plot-based visualizations but faster using color-coded visualizations. The plot-based visualizations improve accuracy for analytical tasks as participants tend to perceive higher discriminability in the data with the \textit{position on a common scale} encoding compared to \textit{color}---interestingly, this result is aligned with perceptual accuracy rankings previously found in controlled experiments for 2D information visualizations \cite{cleveland1984, munzner2014}. Our findings also reveal that plot-based visualizations more often preserve user performance compared to color-coded visualizations with increasing task complexity.} Based on our findings, we were able to derive a series of takeaways that offer practical design guidelines for visualization researchers and practitioners. \highlight{Lastly, we presented a qualitative evaluation with four practitioners from the 3D urban domain. The interviews confirmed the overall usefulness of the visualizations for 3D spatiotemporal urban data analysis, and revealed practitioners’ impressions and suggestions for further work.}

\noindent \textbf{Limitations and future work.} We acknowledge some limitations of our study. Although our quantitative study included participants with considerable experience with 3D visualizations, \highlight{none of them were practitioners working with 3D urban visualizations in a professional setting. We also acknowledge that the findings of our study may be limited by a relatively small sample of participants---mostly graduate students. We feel that general trends found in our study would still apply across domain practitioners as they are familiar with 3D visualizations, similarly to the recruited participants. However, further work is needed to verify how our study results alter with a large sample and how well the visualizations fare when used by domain practitioners.}
%
%
%
Additionally, we recruited participants who might be more familiar with some of the techniques (\textit{e.g.}, temporal juxtaposition) than others (\textit{e.g.}, embedded view). While we believe the general trends still apply, it is possible that this effect could bias the results in favor of some techniques.


\highlight{We also acknowledge the limitations of our visual designs in tackling a key challenge encountered in 3D urban spaces: occlusion. We carefully controlled the amount of occlusion between buildings in our trials as we primarily aimed at evaluating the influence of space and time at different granularities on users' ability to interpret attribute values with different visualizations. Hence, we refrained from adding too many factors to an already complex experimental design. Our discussion section provides thoughts on how (partial) occlusion impacts performance, but further work is needed and could consider a combination of our designs with de-occlusion designs such as the ones presented by Elmqvist \textit{et al.} \cite{elmqvist2008}}.
Another limitation is that we used a single 3D city model in the study, which reflects only a subset of possible geographical configurations. Additional studies may be necessary to measure participants' performance in a more diverse range of model types. While plot-based visualizations performed well for a medium-density model with partial occlusion, they may provide less of a benefit for a high-density model with strong occlusion between buildings.
%
%
\highlight{Our city model also did not include buildings with facade textures that portray structural information.
Further work could consider blending thematic and structural data, and study to what degree users perceive each information individually.}
\highlight{Furthermore, our city model only included buildings with regular geometry. It is possible that irregular geometries could affect the perception of the representation of the thematic variable in our visualizations. Although we believe general trends of our study findings would persist in building models with less pronounced deformations (\textit{e.g.}, facades with smooth, wide curves), further work is needed to consider cases with acute geometric irregularity. In such cases, irregularities may cause undesired distortions on the plot representation of an embedded visualization that  conforms to the surface geometry like ours; therefore, further work could investigate how well the visualization fares with increasing levels of irregularity and consider redesigning it for legibility purposes, if needed. 
Some redesign ideas could build on prior work from GIS research on varying the level-of-detail of building models \cite{biljecki2014}, or consider using proxy surfaces as simplified, smoother approximations of a given polygonal model~\cite{Calderon2017, rocha2018}.

}
%

\highlight{Finally, although we covered a spectrum of analytical tasks elicited in our task characterization with domain experts, our study was also limited to a few tasks---\task{spatial search} for a \task{facade region} with \task{extreme} attribute value over a time period (a \task{time step} or a \task{time interval}).
It is possible that other task settings could have yielded different study results and may be worth further investigating---\textit{e.g.}, in \task{temporal comparison} tasks, where target locations are already known by users, it is very likely that time performance differences between plot-based and color-coded visualizations would be less pronounced. 
Additionally, due to the study feasibility purposes, we limited our tasks to deal with only a few spatial and temporal instances. Although our results provide insights on how well visualizations scale in such cases, future work could concentrate on perceptual scalability and investigate finer granularity levels.

}


%




\section*{Acknowledgements}

\noindent We would like to thank the anonymous reviewers for their constructive comments and feedback.
We acknowledge the support of the CGS-D Program [CGSD3-534866-2019] and Discovery grants [RGPIN/05320-2019, RGPIN/05303-2019] from NSERC,  CNPq [316963/2021-6], and FAPERJ [E-26/202.915/2019, E-26/211.134/2019].


\bibliographystyle{abbrv-doi}

\bibliography{template}
\end{document}